# Extracting a Discriminative Structural Sub-Network for ASD Screening using the Evolutionary Algorithm

M. Amin, F. Safaei, N. S. Ghaderian

Faculty of Computer Science and Engineering, Shahid Beheshti University G.C., Evin 1983963113, Tehran, IRAN

moh.amin@mail.sbu.ac.ir, f_safaei@sbu.ac.ir, n.ghaderian@mail.sbu.ac.ir

**Abstract**

Autism spectrum disorder (ASD) is one of the most significant neurological disorders that disrupt a person's social communication skills. The progression and development of neuroimaging technologies has made structural network construction of brain regions possible. In this paper, after finding the discriminative sub-network using the evolutionary algorithm, the simple features of the sub-network lead us to diagnose autism in various subjects with plausible accuracy (76% on average). This method yields substantially better results compared to previous researches. Thus, this method may be used as an accurate assistance in autism screening.

## 1. Introduction

Neurological diseases are either biochemical or electrical based and they cause specific symptoms in the patient. Alzheimer, Parkinson, Multiple Sclerosis, Autism Spectrum Disorder (ASD) are the most important examples of neurological diseases [5].

Autism spectrum disorder (ASD) is one of the most significant neurological disorders that disrupts a person's social communication skills. The progression and development of neuroimaging technologies and modeling the connections between various brain regions using graphs, has led to screening these disorders with higher speed and accuracy.

Complex networks consist of nodes that are related to each other and have specific features that differentiates them from random networks. The reason for this complexity is that the behavior of the network cannot be derived from the behavior of a specific nodes. The internet, social networks, neural or brain region networks, the relation between proteins, airlines, VLSI circuits etc. are just a few examples of complex networks. According to recent studies, the brain is a modular network that follows the features of the small-world [1].

Using neuroimaging technologies and methods it is possible to derive functional and structural networks of the brain. DTI is a neuroimaging technique based on MRI that allows us to estimate the position, direction and anisotropy of the brain's white matter tracts. Structural brain networks are made using image processing done on DTI that more details are discussed in [1]. In structural brain networks the nodes are usually brain regions and the weight of the edges are the number of fibers between these regions.

Various studies have been conducted on structural brain networks and many of them are focused on the classification of these networks; for example the classification of these networks into two classes of healthy control and Autism spectrum disorder.

The classification of complex networks has many usages i.e. time series anomaly detection [2], link prediction [2,3], cancer cells detection [4], recognition of type of proteins [4], classification in graph based search engines [4], modeling large networks [4], comparing social networks [5], diagnosing neurological disorders [5] etc.

There have been many proposed classification methods for networks that mainly fall into either one of two categories, kernel based or feature based [2, 4, 6-8]. Kernel based methods are very slow because of algorithm complexity [2, 9].

Most previous studies on network classification have used complete networks for their classification. There are logical reasons to choose sub-networks to classify instead of complete networks. A number of them are as follows:

1. When using complete network we usually use the edges as features that causes over-fit because of too many features and too little scans.
2. Specific sub-networks of the brain can give us more info on different disorders and lower the risk of neutralizing the features of a disorder via some nodes.
3. Processing speed increases faster while processing sub-networks.
4. Better interpretation can be achieved when we have a discriminative sub-network.

The aforementioned sub-networks are not specific regions in regards to functionality or structure, but a subdivision of nodes and edges that have been chosen from the main network [10].

In this study, a method has been proposed that uses an evolutionary algorithm to find the discriminative sub-network in the classification, and by utilizing theory of graph, metrics extracts and completes the classification. The chosen sub-network has fewer edges and nodes than the complete network and this sub-network has gained high scores in evolutionary algorithm evaluation thus, the classification is completed faster and more accurately than previous research studies.

In Section 2 previous studies are described and in Section 3, the proposed method has been explained. In Section 4 the results of the simulation and in Section 5 the results of the study and future work are demonstrated.

## 2. Related Literature

There have been many studies conducted on neurocognitive disorders. With the development of technology and machine learning and deep learning methods, researchers are aiming towards feature-based methods. The basis of these methods is the extraction of structural brain features and utilizing machine learning models to diagnose disorders.

Basset and Sporns [11] were the first to suggest neurological networking. They showed that in the era of big data, neurology needs statistical inference and other theoretical ideas to be able to search the structure and functionality of the brain.

Crossly et al. [12] showed that the placement of hubs can change parallel to each other at high rates and this change can transform the type and steps of cognitive processing.

Dennis and Thompson [13] studied the possibility that dementia can be slowed down or stopped and tried to derive the first symptoms of Alzheimer using a brain network.

Rubinov and Sporns [14] presented a group of local and global features for brain networks that can be used to train a machine learning model.

Dodero et al. [15] attempted to classify brain networks using kernel based methods.

Tolan and Isik [5] proposed a binary classification model that differentiates between a person with autism and a person with healthy controls. They calculated the local and global features of the structural and functional brain networks and used this feature vector to train different machine learning models. In order to aggregate the local features they used minimum, maximum, standard deviation and etc methods. In addition to these features, the gender of each person is also considered as an item in the feature vector. They used the relief method [17] in order to decrease the feature vector dimensions and choose the best features. To calculate feature values, they utilized BRAPH tools [16].

They used the LOOCV method as an evaluation strategy and apply their method on the data in [18] for training and testing machine learning models.

Petrov et al. [19] proposed a method that can differentiate between a healthy control and autism brain. They used 16 features that are proportionate to the value of the edges and network topology features. Eventually, they utilized the SVM model for classification.

For choosing efficient sub-networks for classification, Brown et al. [10] proposed an efficient method. They presented an optimization problem that specifies which nodes and edges to keep and which ones to put aside.

The aforementioned methods have disadvantages; the first and foremost being the low accuracy. Kernel-based and some feature-based methods are also slow. The proposed method in this paper has higher accuracy and speed when compared to previous methods as instead of processing the whole network; it only processes the discriminative sub-network.

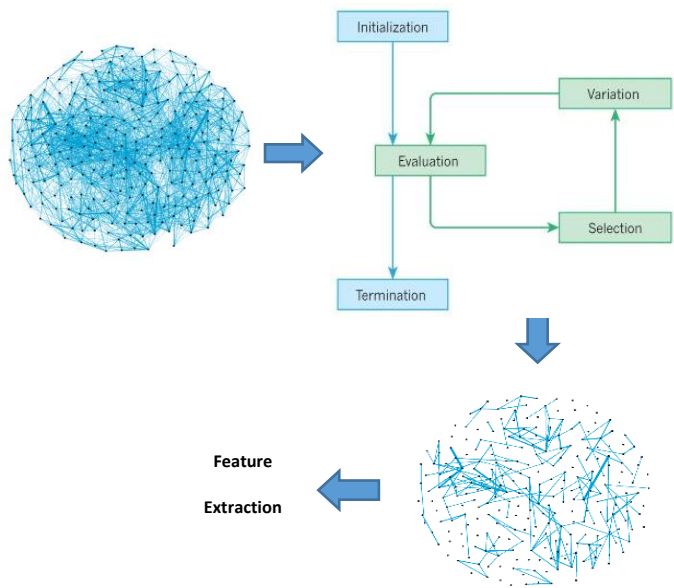

**Figure 1.** Illustration the proposed method. The structural brain sub-network was derived after the evolutionary algorithm was run, and was used for classification.

In this paper, for the first time, a method has been proposed comprehensively that uses the evolutionary algorithm to choose an effective sub-network and using network topology features, classification is done with sufficient speed and accuracy. In the following section, the proposed method is extensively described.

## 3. The Proposed Method

With the vast development of neuroimaging technology, deriving the functional and structural networks of different brain regions has been made possible. In this article, a novel method has been proposed that uses the evolutionary algorithm to find an efficient discriminative sub-network, and classify the networks into two groups of healthy control and autism spectrum disorder via extracting the topology features. In this article, the dataset in [18] has been used. The information is presented in table 1.

Table 1. Dataset Information

| HC | ASD | Total |
|---|---|---|
| 42 | 51 | 93 |

For implementing the proposed method, the programming language python version 2.7, the complex network library NetworkX version 2.1, the machine learning library scikit-learn version 1.0.9.2 and the editor PyCharm 2018 version were utilized.

The main process of the proposed method can be seen in Figure 1. The first step is to find the discriminative sub-network; a sub-network in which the metrics calculated have a greater differentiation between the healthy control and autism spectrum disorder groups and thus, classification is more accurate. Each structural network of the dataset has 264 nodes (each node is a region of the brain) therefore, each node has a label with a value between 0 and 255.

Assume a string of length 264 in which each unit can be zero or one. The position of each unit corresponds to the label of each node. If the unit in that position is one, the node exists in the network. If the unit is zero, the node doesn't exist in the network (and neither does the edges leading up to the node.). In the case that all units in the string are one, all nodes exist thus the network is complete.

## 3.1 Evolutionary Algorithm

Our problem is finding the sub-network that yields a higher level of differentiation between the network topology metrics of the healthy control and autism groups. As a result, this sub-network also yields a more accurate classification. According to the number of nodes, if all possible states are examined, there would be $2^{264}-2$ states to examine. Therefore, evolutionary algorithms are used to find the discriminative sub-network.

*Presentation method*

One of the important beginning steps for using evolutionary algorithms is finding a suitable presentation method for the solutions. Choosing the suitable presentation method facilitates the process of mapping the genotypes to the phenotypes. As mentioned before, the presentation method for the solutions is in binary.

*Initial Population*

For launching the evolutionary algorithm, an initial population is essential. The population must have a high diversity and the number of people in the population must be suitable. If not, the algorithm's execution speed will decrease. There are two methods to choose an initial population, random population and heuristic population. In this article, random population is used. The initial population has 80 solutions that have been picked at random.

*Fitness function*

To calculate the fitness of each solution, the threshold is applied to each network. Then, the nodes (and edges connected to them) corresponding to the zero units in the solution are deleted. Thereupon, using the features described in the following paragraphs, a decision tree is trained. Finally, evaluation is done via the LOOCV evaluation method. The average classification accuracy is the value of the solution fitness.

*Selection*

The best solution of a generation is selected via the truncated based selection method. This method sorts the solutions based on fitness and selects $p$ percent of them. Thereupon, each selected solution is reproduced $1/p$ times.

*Cross-over*

The cross-over operator is similar to biological rebirth and synthesis. There are various synthesis methods. In this paper, the one point cross-over method is used in which two solutions are selected first then, a point between 1 and the length of the solution is randomly chosen. Finally, the two parts, before and after the chosen point are switched.

*Stop Condition*

Choosing a rational ending condition is essential. If not chosen, the algorithm will never end. In this study, there are two stop conditions. Whichever one is reached, the algorithm stops. The first one is the number of generations constraint and the other is reaching the desired accuracy limit.

## 3.2 Features Selection for Classification

For each network, a specific edge weight threshold must be applied first. This threshold has been chosen according to [1].

The threshold for each network is a value with which the network sparseness value becomes 0.025. This value is demonstrated by $S(G)$ for network $G$ with edge set $E$, and node set $V$. $S(G)$ is expressed by the following equation.

$$S(G) = \frac{2|E|}{n(n-1)} \qquad (1)$$

In which, $n$ is the number of nodes in the network.

The specified features for each network are categorized as two groups, local and global.

The global network features for network $G$ with an edge set $E$, and a node set $V$ are as follows.

1. The transitivity feature that is expressed as follows

$$T(G) = \frac{3 \times ltr}{lt} \qquad (2)$$

In which $ltr$ is the number of all possible triangles in network $G$ and $lt$ is the number of network triads.

2. The average clustering feature that is given by

$$C(G) = \frac{1}{n}\sum_{i=1}^{n} C_i \qquad (3)$$

In which $C_i$ is the node clustering coefficient and n is the number of nodes in the network.

3. The *global efficiency* feature that is calculated as

$$E(G) = \frac{1}{n}\sum_{i \in V} \frac{\sum_{j \in V, j \neq i} \frac{1}{d_{ij}}}{n-1} \qquad (4)$$

In which, $d_{ij}$ is the shortest route between two nodes, $i$ and $j$. Moreover, $n$ is the number of nodes in the network. Local features are also determined for each node. The following included:

1. The closeness feature that is calculated as

$$CC(G) = \frac{n-1}{\sum_{v=1}^{n-1} d(u,v)} \quad (5)$$

In which $d(u,v)$ is the shortest route between nodes $u$ and $v$; $n$ is the number of nodes in the network.

2. The betweenness feature that is determined by the following equation

$$BC(G) = \sum_{s,t \in V} \frac{\sigma(s,t|v)}{\sigma(s,t)} \quad (6)$$

In which $\sigma(s,t)$ is the number of short routes between $s$ and $t$, and $\sigma(s,t|v)$ is the number of routes that include node $v$.

3. The triangle feature that derives the number of triangles for node $v$ that include itself.

For the local features, three aggregation methods (maximum, average and standard deviation) are used. Therefore, there are 12 features for each network.

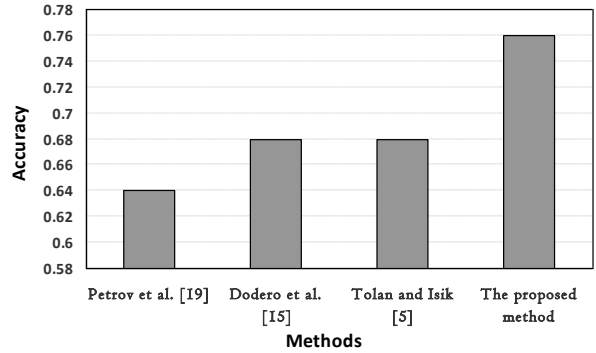

**Figure 2.** The mean accuracy comparison between the proposed method and the related work

Table 2. LOOCV Evaluation Report

| | precision | Recall | f1-score |
|---|---|---|---|
| Discriminative sub-network | | | |
| Healthy Control | **0.74** | **0.71** | **0.72** |
| Autism Dysfunction | **0.77** | **0.80** | **0.79** |
| Accuracy | - | - | **0.76** |
| Macro Average | **0.76** | **0.76** | **0.76** |
| Weighted Average | **0.76** | **0.76** | **0.76** |
| Complete network | | | |
| | precision | recall | f1-score |
| Healthy Control | 0.37 | 0.44 | 0.40 |
| Autism Dysfunction | 0.47 | 0.39 | 0.43 |
| Accuracy | - | - | 0.41 |
| Macro Average | 0.42 | 0.42 | 0.41 |
| Weighted Average | 0.42 | 0.41 | 0.41 |

## 4. Simulation Results

In the previous section, we discussed the method proposed in order to find the discriminative sub-network using the evolutionary algorithm. The derived sub-network has 127 nodes. From here the sub-network may be used instead of the complete network for classification.

The evaluation of choosing a sub-network and its effect on classification accuracy was done via the LOOCV method. This procedure was done 100 times using the extracted sub-network. The results have been reported in Table 2. The 12 features mentioned were used in training the decision tree. As depicted in table 2, the important sub-network is effective and the recall, precision and accuracy are much higher than that of a complete network. Many researchers have studied structural brain networks and various methods of classification into the two groups of healthy control and autism have been proposed and implemented, Figure 2 shows the accuracy comparison between these methods and the one proposed in this paper.

## 5. Conclusions

Autism spectrum disorder (ASD) is one of the most significant neurological disorders that disrupts a person's social communication skills. Developing new screening methods using structural and functional brain networks is valuable as diagnosing autism with plausible accuracy at a young age, while symptoms are not yet visible, can be beneficial to the patient. In this paper it is assumed that instead of processing the whole brain structural network, it is possible to find a sub-network which depicts the difference between the structural network of a healthy brain and a brain with autism more clearly. The sub-network extraction (instead of whole brain network) must also increase the speed and accuracy of classification. In future studies, more features may be used, various models of machine learning can be experimented and other types of evolutionary algorithms can be used.

## References

[1] J. D. Rudie et al., "Altered functional and structural brain network organization in autism," NeuroImage: Clinical, vol. 2, pp. 79-94, 2013.

[2] S. Bonner, J. Brennan, G. Theodoropoulos, I. Kureshi, and A. S. McGough, "Deep topology classification: A new approach for massive graph classification," in 2016 IEEE International Conference on Big Data (Big Data), 2016.

[3] A. Pecli, M. C. Cavalcanti, and R. Goldschmidt, "Automatic feature selection for supervised learning in link prediction applications: a comparative study," Knowledge and Information Systems, vol. 56, no. 1, pp. 85-121, Oct. 2017.

[4] JP.Canning, EE.Ingram, SN.Wolff, AM.Ortiz, NK.Ahmed, RA.Rossi, KRB.Schemitt, S. Soundarajan," Predicting Graph Categories from Structural Properties,"CoRR,vol. abs/1805.02682, 2018.

[5] E. Tolan and Z. Isik, "Graph Theory Based Classification of Brain Connectivity Network for Autism Spectrum Disorder," in Bioinformatics and Biomedical Engineering, Springer International Publishing, 2018, pp. 520-530.

[6] S.Bonner, J.Brennan, G. Theodoropoulos, I. Kureshi and AS.McGough," Efficient Comparison of Massive Graphs Through The Use Of 'Graph Fingerprints,' "Twelfth Workshop on Mining and Learning with Graphs (MLG) '16 San Francisco, California USA, 2016.

[7] B. Kantarci and V. Labatut, "Classification of Complex Networks Based on Topological Properties," in 2013 International Conference on Cloud and Green Computing, 2013.

[8] A. Grover and J. Leskovec, "node2vec," in Proceedings of the 22nd ACM SIGKDD International Conference on Knowledge Discovery and Data Mining-KDD '16, 2016.

[9] N. S. Ketkar, L. B. Holder, and D. J. Cook, "Empirical comparison of graph classification algorithms," in 2009 IEEE Symposium on Computational Intelligence and Data Mining, 2009.

[10] C. J. Brown, S. P. Miller, B. G. Booth, J. G. Zwicker, R. E. Grunau, A. R. Synnes, V. Chau, and G. Hamarneh, "Predictive connectome subnetwork extraction with anatomical and connectivity priors," Computerized Medical Imaging and Graphics, vol. 71, pp. 67–78, Jan. 2019.

[11] D. S. Bassett and O. Sporns, "Network neuroscience," Nature Neuroscience, vol. 20, no. 3, pp. 353-364, Mar. 2017.

[12] N. A. Crossley et al., "Cognitive relevance of the community structure of the human brain functional coactivation network," Proceedings of the National Academy of Sciences, vol. 110, no. 28, pp. 11583-11588, Jun. 2013.

[13] E. L. Dennis and P. M. Thompson, "Functional Brain Connectivity Using fMRI in Aging and Alzheimer's Disease," Neuropsychology Review, vol. 24, no. 1, pp. 49-62, Feb. 2014.

[14] M. Rubinov and O. Sporns, "Complex network measures of brain connectivity: Uses and interpretations," NeuroImage, vol. 52, no. 3, pp. 1059–1069, Sep. 2010.

[15] L. Dodero, H. Q. Minh, M. S. Biagio, V. Murino, and D. Sona, "Kernel-based classification for brain connectivity graphs on the Riemannian manifold of positive definite matrices," in 2015 IEEE 12th International Symposium on Biomedical Imaging (ISBI), 2015.

[16] M. Mijalkov, E. Kakaei, J. B. Pereira, E. Westman, and G. Volpe, "BRAPH: A graph theory software for the analysis of brain connectivity," PLOS ONE, vol. 12, no. 8, p. e0178798, Aug. 2017.

[17] K. Kira, L. A.Rendell, "The feature selection problem: traditional methods and a new algorithm," Aaai, vol. 2 , 1992.

[18] J. A. Brown, J. D. Rudie, A. Bandrowski, J. D. Van Horn, and S. Y. Bookheimer, "The UCLA multimodal connectivity database: a web-based platform for brain connectivity matrix sharing and analysis," Frontiers in Neuroinformatics, vol. 6, 2012.

[19] D. Petrov,Y. Dodonova, L. Zhukov." Differences in Structural Connectomes between Typically Developing and Autism Groups.",INFORMATION TECHNOLOGIES AND SYSTEMS, 2015, pp. 763777.